\documentclass[a4paper,11pt]{article}%
\usepackage{geometry}
\usepackage{amsmath}
\usepackage{amsfonts}
\usepackage{amssymb}
\usepackage{graphicx}
\usepackage{indentfirst}
\usepackage[small,bf]{caption}
\usepackage{slashed}
\providecommand{\U}[1]{\protect\rule{.1in}{.1in}}

\begin{document}

\date{}
\title{\textbf{Yang-Mills theory in the maximal Abelian gauge in presence of scalar matter fields}}
\author{\textbf{M.~A.~L.~Capri}\thanks{caprimarcio@gmail.com}\,\,,
\textbf{D.~Fiorentini }\thanks{diegofiorentinia@gmail.com}\,\,,
\textbf{S.~P.~Sorella}\thanks{silvio.sorella@gmail.com}\,\,\,,\\[2mm]
{\small \textnormal{  \it Departamento de F\'{\i }sica Te\'{o}rica, Instituto de F\'{\i }sica, UERJ - Universidade do Estado do Rio de Janeiro,}}
 \\ \small \textnormal{ \it Rua S\~{a}o Francisco Xavier 524, 20550-013 Maracan\~{a}, Rio de Janeiro, Brasil}\normalsize}
\maketitle

\begin{abstract}
We address the issue of the all order multiplicative renormalizability of $SU(2)$ Yang-Mills theories quantized in the maximal Abelian gauge in presence of scalar matter fields. The non-linear character of the maximal Abelian gauge requires the introduction of quartic interaction terms in the Faddeev-Popov ghosts, a well known feature of this gauge.  We show that, when scalar matter fields are introduced,  a second quartic interaction term between scalar fields and Faddeev-Popov ghosts naturally arises. A BRST invariant action accounting for those quartic interaction terms is identified and proven to be multiplicative renormalizable to all orders by means of the algebraic renormalization procedure.

\end{abstract}

\section{Introduction} 
Nowadays, the maximal Abelian gauge \cite{'tHooft:1981ht,Kronfeld:1987vd,Kronfeld:1987ri} is widely employed in  order to investigate nonperturbative aspects of Yang-Mills theories. This gauge turns out to be suitable for the study of the dual superconductivity
mechanism for color confinement \cite{scon}, according to which Yang-Mills
theories in the low energy region should be described by an effective
Abelian theory \cite{Ezawa:bf,Suzuki:1989gp,Suzuki:1992gz,Hioki:1991ai,Sakumichi:2014xpa} in
the presence of monopoles. The condensation of these magnetic charges leads
to a dual Meissner effect resulting in quark confinement. In the maximal Abelian gauge, the Abelian
configuration is identified with the diagonal components $A_{\mu }^{3}$ of the gauge field 
corresponding to the diagonal generator of the Cartan subgroup of $SU(2)$. The
remaining off-diagonal components $A_{\mu }^{a}$,
$a=1,2$, corresponding to the off-diagonal generators of $SU(2)$, are expected to acquire a mass
through a dynamical mechanism, thus decoupling at low energies. This phenomenon is known as Abelian 
dominance and is object of intensive investigation, both from analytic and from numerical lattice simulations. 
\newline
\newline
From the analytic side, evidence for the dynamical mass generation for the off-diagonal components of the gauge field can be found in \cite{Schaden:1999ew,Kondo:2001nq,Dudal:2004rx}, while \cite{Amemiya:1998jz,Bornyakov:2003ee,Gongyo:2013sha}
 are devoted to numerical studies.\\\\Besides being a renormalizable gauge   \cite{Min:1985bx,Fazio:2001rm,Gracey:2005vu}, the maximal Abelian gauge enjoys the important property 
of exhibiting a lattice formulation \cite{Amemiya:1998jz,Bornyakov:2003ee,Gongyo:2013sha,Mendes:2006kc,Mihari:2007zz}, a property which allows to compare analytic and numerical results. In particular, this important feature of the maximal Abelian gauge has made possible the study, from the numerical lattice point of view, of the behaviour of the two-point gluon correlation function in the non-pertutbative infrared region, providing evidence for the Abelian dominance as well as for the confining character of the propagator of the  Abelian gluon component \cite{Amemiya:1998jz,Bornyakov:2003ee,Gongyo:2013sha,Mendes:2006kc,Mihari:2007zz}. This issue has also been addressed through analytical methods by taking into account the existence of the Gribov copies  \cite{Gribov:1977wm} which, as in any covariant and renormalizable gauge,  affect the maximal Abelian gauge \cite{Bruckmann:2000xd,Guimaraes:2011sf,Capri:2013vka}. Here, proceeding in a way similar to the Landau gauge \cite{Sobreiro:2005ec,Vandersickel:2012tz}
, a few properties of the so called Gribov region have been derived together with the restriction of the domain of integration in the functional integral to the Gribov horizon, see for instance refs.\cite{Capri:2006vv,Capri:2006cz,Capri:2008ak,Capri:2008vk,Capri:2010an} for the details of the Gribov issue on the maximal Abelian gauge.  Remarkably, the agreement between the lattice numerical results and the analytic calculations based on the restriction to the Gribov region looks quite good \cite{Mendes:2006kc,Capri:2008ak}, confirming the expectation that the study of the Gribov problem is of great relevance for gluon confinement.\\\\Nevertheless, so far, the study of the correlation function in the maximal Abelian gauge has been done only for the gluon sector, without including matter fields, {\it i.e.} spinor and scalar fields. To our knowledge, unlike the Landau gauge, no available non-perturbative studies of the two-point matter correlation functions are available in the maximal Abelian gauge, and this from both analytical and numerical simulations. \\\\This work aims at starting an analytic study of the non-perturbative behaviour of the correlation functions for matter fields in the maximal Abelian gauge, along the lines recently outlined in the case of the Landau gauge \cite{Dudal:2008sp,Dudal:2011gd,Capri:2014bsa}, where it has been possible to recover the behaviour of the propagators for scalar and spinor fields observed in lattice simulations \cite{Maas:2011yx,Maas:2010nc,Furui:2006ks,Parappilly:2005ei} from an analytic point of view \cite{Capri:2014bsa}. This study might be of relevenace for several reasons as, for instance: investigate to what extent the Abelian dominance affects the matter sector, make prediction for the propagator of scalars and quark fields which might be compared with lattice numerical simulations, study of the confining character of the correlation functions. \\\\As first step in this endeavour, we need to establish the all orders multiplicative renormalizability of the maximal Abelian gauge in presence of matter fields, a topic which, till now, has not yet been addressed.  This is the goal of the present paper. Although the renormalizability of the maximal Abelian gauge in presence of the matter fields is an expected feature, we shall see that it is not a straightforward matter, requiring in fact a  nontrivial analysis. This is due to the non-linear character of the maximal Abelian gauge which gives rise to a rather complex Faddeev-Popov operator. It was already pointed out that the structure of this operator requires the introduction of a quartic interaction between ghosts \cite{Min:1985bx,Fazio:2001rm,Gracey:2005vu}. Only at the very end of the whole renormalization process the gauge parameter entering the quartic interaction can be set to zero \cite{Min:1985bx,Fazio:2001rm,Gracey:2005vu}, thus recovering the genuine  maximal Abelian gauge condition. In this work, we shall see that this feature generalises to the case of scalar matter fields, {\it i.e.} a quartic interaction between scalar fields and Faddeev-Popov ghosts naturally arises due to the non-linearity of the gauge condition. As a consequence, a second gauge parameter associated to this new term has to be introduced. As in the case of the quartic ghost term, this second gauge parameter can be set to zee only at the very end of the renormalization process. \\\\The present work is organised as follows. In Sect.2 we briefly discuss the maximal Abelian gauge and the corresponding gauge fixing. In Sect.3 we elaborate on the quartic interactions required to renormalize the theory. Sect.4 is devoted to establish the set of Ward identities needed for the all orders proof of the renormalizability. In Sect.5 we present the algebraic characterisation of the most general invariant local counterterm, establishing the all orders multiplicative renormalizability of the theory. Sect.6 collects our conclusion.

\section{Quantizing gauge theories in the maximal Abelian gauge}

In order to introduce the maximal Abelian gauge, we start by considering a Lie algebra valued gauge field $\mathcal{A}_\mu $  for the gauge
group $SU(2),$ whose generators $T^A\,\;(A=1,..,3)\,
$ \begin{equation}
\left[ T^A,T^B\right] =\varepsilon ^{ABC}T^C\,\,  \label{la}
\end{equation}
are chosen to be antihermitean and to obey the orthonormality condition $\,
\mathrm{Tr}\left( T^AT^B\right) =\delta ^{AB}$. 
Following  \cite{'tHooft:1981ht,Kronfeld:1987vd,Kronfeld:1987ri} we decompose  $\mathcal{A}_\mu $  into off-diagonal and diagonal components 
\begin{equation}
\mathcal{A}_\mu =\mathcal{A}_\mu ^AT^A=A_\mu ^aT^a+A_\mu T^{\,3},  \label{cd}
\end{equation}
where $a=1,2$ and $T^3$ is the diagonal generator of the Cartan subgroup sf $SU(2)$. Analogously, decomposing the field strength, we obtain
\begin{equation}
\mathcal{F}_{\mu \nu }=\mathcal{F}_{\mu \nu }^AT^A=F_{\mu \nu }^aT^a+F_{\mu
\nu }T^{\,3},  \label{fs}
\end{equation}
with the off-diagonal and diagonal components given, respectively, by
\begin{eqnarray}
F_{\mu \nu }^a &=&D_\mu ^{ab}A_\nu ^b-D_\nu ^{ab}A_\mu ^b\,,  \nonumber \\
F_{\mu \nu } &=&\partial _\mu A_\nu -\partial _\nu A_\mu +g \varepsilon
^{ab}A_\mu ^aA_\nu ^b\,,  \label{fscomp}
\end{eqnarray}
where the covariant derivative $D_\mu ^{ab}$ is defined with respect to the
diagonal component $A_\mu $
\begin{equation}
D_\mu ^{ab}\equiv \partial _\mu \delta ^{ab}-g\varepsilon ^{ab}A_\mu
\,\,\,\,\,\,,\,\,\,\varepsilon ^{ab}\equiv \varepsilon ^{ab3}\,\,\,.
\label{cder}
\end{equation}
For the  classical gauge invariant starting action, we have 
\begin{equation}
S_{cl} = S_{YM} + S_{matter}    \;, \label{scl}
\end{equation} 
where $S_{YM}$ stands for the Yang-Mills action 
\begin{equation}
S_{YM}=\int d^4x\; \frac{1}{4} \left(F_{\mu\nu}^aF_{\mu\nu}^a+F_{\mu\nu}F_{\mu\nu}\right)  \;, \label{ym}
\end{equation}
while $S_{matter}$ denotes the  action of real scalar matter fields in the adjoint representation of the gauge group $SU(2)$, namely 
\begin{eqnarray}
S_{matter}&=&\int d^4x \left(\frac{1}{2}(D_{\mu}^{AB}\phi ^B)^2+\frac{m^2_{\phi}}{2}\phi^A\phi^A+\frac{\lambda}{4!}(\phi ^A\phi ^A)^2\right)\nonumber\\
&=&\int d^4x \left\{(\partial_{\mu}\phi^a)(\partial_{\mu}\phi^a)+(\partial_{\mu}\phi)(\partial_{\mu}\phi)
-2g^2\varepsilon^{ab}\left[(\partial_{\mu}\phi)\phi^a A_{\mu}^b
-(\partial_{\mu}\phi^a)\phi A_{\mu}^b+(\partial_{\mu}\phi^a)\phi^b A_{\mu}\right]
\right.\nonumber\\
&&+g^2\left[A_{\mu}^aA_{\mu}^a\left(\phi^b\phi^b+\phi\phi\right)
+A_{\mu}A_{\mu}\phi^a\phi^a -A_{\mu}^aA_{\mu}^b\phi^a\phi^b
-2A_{\mu}^aA_{\mu}\phi^a\phi\right]
\nonumber\\
&&\left.+\frac{m^2_{\phi}}{2}\left(\phi^a\phi^a+\phi\phi\right)
+\frac{\lambda}{4!}\left[\left(\phi^a\phi^a\right)^2+2\phi^2\phi^a\phi^a+\phi^4\right]\right\}      \;,
\label{smatter}
\end{eqnarray}
where, as in eq.\eqref{cd}, the scalar field $\mathcal{\phi}= \phi^A T^A$ is decomposed into off-diagonal and diagonal components, {\it i.e.} 
\begin{equation}
\phi^AT^A=\phi^aT^a+ \phi T^{\,3}.  \label{sf}
\end{equation}
The classical action (\ref{scl}) is left invariant
by the gauge transformations
\begin{eqnarray}
\delta A_{\mu }^{a} &=&-D_{\mu }^{ab}{\omega }^{b}-g\varepsilon ^{ab}A_{\mu
}^{b}\omega \;,  \nonumber \\
\delta A_{\mu } &=&-\partial _{\mu }{\omega }-g\varepsilon ^{ab}A_{\mu
}^{a}\omega ^{b}\;,  \label{gauge}
\end{eqnarray}
and 
\begin{equation}
\delta \phi^{a}=g\varepsilon^{ab}\phi\,\omega ^{b}-g\varepsilon ^{ab}\phi ^b \omega \,,\qquad \delta \phi=-g\varepsilon^{ab}\phi^a\omega ^b  \label{sct} \;. 
\end{equation} 
The maximal Abelian gauge condition amounts to impose that the off-diagonal
components $A_{\mu }^{a}$ of the gauge field obey the following nonlinear condition
\begin{equation}
D_{\mu }^{ab}A_{\mu }^{b}=0\;,  \label{offgauge}
\end{equation}
which follows by requiring that the auxiliary functional
\begin{equation}
\mathcal{R}[A]=\int {d^{4}x}A_{\mu }^{a}A_{\mu }^{a}\;,  \label{fmag}
\end{equation}
is stationary with respect to the gauge transformations (\ref{gauge}).
Moreover, as it is apparent from the presence of the covariant derivative $%
D_{\mu }^{ab}$, equation (\ref{offgauge}) allows for a residual local $U(1)$
invariance corresponding to the diagonal subgroup of $SU(2)$. This
additional invariance has to be fixed by means of a further gauge condition
on the diagonal component $A_{\mu }$, which is usually chosen to be of the
Landau type, namely
\begin{equation}
\partial _{\mu }A_{\mu }=0\;.  \label{dgauge}
\end{equation}
The Faddeev-Popov operator, $\mathcal{M}^{ab}$, corresponding to the gauge
condition (\ref{offgauge}) is easily derived by taking the second variation
of the auxiliary functional $\mathcal{R}[A]$, being given by
\begin{equation}
\mathcal{M}^{ab}=-D_{\mu }^{ac}D_{\mu }^{cb}-g^{2}\varepsilon
^{ac}\varepsilon ^{bd}A_{\mu }^{c}A_{\mu }^{d}\;.  \label{offop}
\end{equation}
It enjoys the property of being Hermitian and, as pointed out in \cite
{Bruckmann:2000xd}, is the difference of two positive semidefinite operators
given, respectively, by $-D_{\mu }^{ac}D_{\mu }^{cb}$ and $g^{2}\varepsilon
^{ac}\varepsilon ^{bd}A_{\mu }^{c}A_{\mu }^{d}$. \newline
\newline
It is worth to point out that the operator $\mathcal{M}^{ab}$ is non-linear in the gauge fields, a feature which has 
nontrivial consequences in the renormalization process.

\section{BRST symmetry and emergency of quartic interaction terms} 
In order to construct the Faddeev-Popov action corresponding to the gauge conditions \eqref{offgauge},\eqref{dgauge}, we proceed by introducing the nilpotent BRST transformations 
\begin{eqnarray}
sA^{a}_{\mu}& = & -(D^{ab}_{\mu}c^{b}+g\varepsilon^{ab}A_{\mu}^bc)\,,\qquad sA_{\mu}=-(\partial_{\mu}c+g\varepsilon^{ab}A_{\mu}^ac^b)\nonumber\\ 
sc^{a}& = & g\varepsilon^{ab}c^{b}c\,,\qquad sc=\frac{g}{2}\varepsilon^{ab}c^{a}c^b \,, \nonumber \\
s\bar{c}^{a}& = & b^{a}\,,\qquad s\bar{c}=b\,,\qquad
sb^{a}=sb=0\,,\\
s\phi^{a} & = &g \varepsilon^{ab}\phi\,c^{b}-g\varepsilon ^{ab}\phi ^bc\,,\qquad s\phi=-g\varepsilon^{ab}\phi^ac^b\nonumber   \;, 
\label{stransf}
\end{eqnarray}
where $({\bar c}^a, {\bar c}, c^a, c)$ are the Faddeev-Popov ghosts and $(b^a,b)$ are the Nakanishi-Lautrup fields. Further, we introduce the 
$s$-exact gauge fixing term
\begin{eqnarray}
S_{MAG}&=&s\int d^4x\left\{\bar{c}^aD^{ab}_{\mu}A^{b}_{\mu}+\bar{c}A_{\mu}\right\}\nonumber\\
&=&\int d^4x\left\{b^aD_{\mu}^{ab}A_{\mu}^b-\bar{c}^a\mathcal{M}^{ab}c^b
+g\varepsilon^{ab}\bar{c}^acD_{\mu}^{bc}A_{\mu}^c+b\partial_{\mu}A_{\mu}
+\bar{c}\,\partial_{\mu}\left(\partial_{\mu}c+g\varepsilon^{ab}A_{\mu}^ac^b\right)\right\}  \;, \label{smag}
\end{eqnarray}
where $\mathcal{M}^{ab}$ stands for the Faddeev-Popov operator \eqref{offop}. 
Evidently, the gauge-fixed action 
\begin{equation}
S_{cl} + S_{MAG}  \;, \label{act1}
\end{equation}
with $S_{cl}$ given in eq.\eqref{scl}, turns out to be BRST invariant. The action \eqref{act1} is the gauge-fixed action obtained from the BRST construction, usually taken as the starting action in order to evaluate the quantum corrections arising in the renormalization process. However, in the present case, expression \eqref{act1} has to be supplement by the introduction of further quartic terms which originate from the non-linearity of the Faddeev-Popov operator $\mathcal{M}^{ab}$, eq.\eqref{offop}. In fact, as one can observe from expression \eqref{smag}, the interaction term $g^{2} \bar{c}^a \varepsilon^{ac}\varepsilon^{bd}A_{\mu }^{c}A_{\mu }^{d} c^b $ gives rise to divergent Feynman diagrams with four external Faddeev-Popov legs, as one immediatetely realises already at one-loop level by considering the divergent 1PI diagram with four external Faddeev-Popov ghosts  and two internal off-diagonal gauge lines. As already pointed out in \cite{Min:1985bx,Fazio:2001rm,Gracey:2005vu}, such diagrams give rise to counterterms in the Faddeev-Popov ghosts which are not contained in the action \eqref{act1}. Such additional divergences can be taken into account by introducing the following BRST exact terms \cite{Min:1985bx,Fazio:2001rm,Gracey:2005vu}
\begin{eqnarray}
S_{\alpha}&=&s\int d^4\left\{\frac{\alpha}{2}\left(\bar{c}^ab^a
-2\varepsilon^{ab}\bar{c}^a\bar{c}^bc\right) \right.\nonumber\\
&=&\frac{\alpha}{2}\int d^4x\left\{b^ab^a-2g\varepsilon^{ab}b^a\bar{c}^bc
+g^2\bar{c}^a\bar{c}^bc^ac^b\right\}   \;, \label{alfa}
\end{eqnarray}  
where $\alpha$ stands for a suitable gauge parameter. As one can easily figure out, the quartic divergent terms originating from the action  \eqref{act1} can now be reabsorbed in the renormalization of the gauge parameter $\alpha$. \\\\Nevertheless, the term \eqref{alfa} is not the unique new quartic interaction present in the theory when scalar matter fields are added. In fact, it turns out that, due to the presence of the interaction vertices $(\phi\phi AA)$ and $(\phi  (\partial \phi) A) $, a novel quartic term between scalar fields and Faddeev-Popov ghosts, {\it i.e.} $(\phi \phi {\bar c} c)$, is generated at the quantum level.   For example, the 1PI one-loop diagram with two external $\phi$-legs and two external ghost legs connected by two internal gluon lines is logarithmic divergent, giving rise to a quartic divergent term precisely  of the kind of  $(\phi \phi {\bar c} c)$. Once again, such divergent terms are not contained in the  action \eqref{act1}. As such, they would be not re-absorvables. We see therefore that the, due to the nonlinearity of the gauge condition, eq.\eqref{offgauge},  and of the Faddeev-Popov operator, eq.\eqref{offop}, a second quartic terms is needed for renormalizability. In the present case, this novel term is accounted for  by introducing the following exact BRST expression 
\begin{eqnarray}
S_{\beta}&=&s\int d^4\left\{\frac{\beta}{2}\varepsilon^{ab}\phi\phi^a\bar{c}^b\right\} \nonumber\\
&=&\frac{\beta}{2}\int d^4x\left\{g\phi^a\phi^a c^b\bar{c}^b+g\phi^a\phi^bc^a\bar{c}^b
+\phi\phi^a\left(\varepsilon^{ab}b^b-gc\bar{c}^a\right)+g\phi\phi c^a\bar{c}^a\right\}  \;, \label{sbeta}
\end{eqnarray}
where $\beta$ stands for a second gauge parameter. The emergency of divergent terms of the type $(\phi \phi {\bar c} c)$ is now taken into account by an appropriate renormalization of the second gauge parameter $\beta$. 
In conclusion, taking into account the emergency of quartic interaction terms, for the starting gauge-fixed Faddeev-Popov action we have 
\begin{equation}
S = S_{cl} + S_{MAG} +S_{\alpha}+ S_{\beta}  \;.  \label{act2}
\end{equation}  
Looking ate the equations of motion of the field $b^a$, namely 
\begin{equation}
\frac{\delta S}{\delta b^a}=D^{ab}_{\mu}A^b_{\mu}+\alpha\left(b^a-g\varepsilon^{ab\bar{c}^b}\right)
+\frac{\beta}{2}g\varepsilon^{ba}\phi\phi^b
\end{equation}
we see that the original maximal Abelian gauge condition \eqref{offgauge} is recovered in the limit $\alpha,\beta\rightarrow0$. However, as argued before, such limit has to be taken at the very end of the whole renormalization process.  Having identified a suitable starting action, eq.\eqref{act2}, it remains to prove that it is multiplicative renormalizable to all order, a task which we shall face in the following sections by making use of the algebraic renormalization \cite{Piguet:1995er}.

\section{Ward identities} 

Having identified a suitable gauge-fixed action, eq.\eqref{act2}, we proceed to write down the set of Ward identities which we 
shall employ in the proof of the all orders multiplicative renormalizability of expression \eqref{act2}. To that end, following the algebraic renormalization procedure  \cite{Piguet:1995er}, we need to introduce a set of BRST invariant external sources $(\Omega^a_\mu, \Omega_\mu, L^a, L, F^a, F)$ coupled to the non-linear BRST variations of the fields $(A^a_\mu, A_\mu, c^a, c, \phi^a, \phi)$, eqs.\eqref{stransf}, namely 
  
\begin{eqnarray}
S_{ext}&=&\int d^4x\left\{\Omega_{\mu}^a(sA_{\mu}^a)+\Omega_{\mu}(sA_{\mu})+L^a(sc^a)
+L(sc)+F^a(s\phi^a)+F(s\phi)\right\}\nonumber\\
&=&\int d^4x \left\{\Omega_{\mu}^a\left(-D^{ab}_{\mu}c^{b}-g\varepsilon^{ab}A_{\mu}^bc\right)
+\Omega_{\mu}\left(-\partial_{\mu}c-g\varepsilon^{ab}A_{\mu}^ac^b\right)
+g\varepsilon^{ab}L^ac^{b}c
 \right.\nonumber\\
&&\left. +\frac{g}{2}\varepsilon^{ab}Lc^{a}c^b
+g\varepsilon ^{ab}F^a\left(\phi\,c^{b}-\phi ^bc\right)-g\varepsilon^{ab}F\phi^ac^b\right\}  \;,
\end{eqnarray}
with
\begin{equation}
s\Omega_{\mu}^a=s\Omega_{\mu}=sF^a=sF=sL^a=sL=0   \;.
\end{equation}
Therefore, for the complete BRST invariant starting action $\Sigma$, we get 
\begin{eqnarray}
\Sigma &=&S_{YM} + S_{matter}+S_{MAG}+S_{\alpha}+S_{\beta} + S_{ext} \nonumber\\
&=&\int d^4x\left\{ \frac{1}{4}\left(F_{\mu\nu}^aF_{\mu\nu}^a+F_{\mu\nu}F_{\mu\nu}\right)
+b^aD_{\mu}^{ab}A_{\mu}^b-\bar{c}^a\mathcal{M}^{ab}c^b
+g\varepsilon^{ab}\bar{c}^acD_{\mu}^{bc}A_{\mu}^c+b\partial_{\mu}A_{\mu}
\right.\nonumber\\
&&+\bar{c}\,\partial_{\mu}\left(\partial_{\mu}c+g\varepsilon^{ab}A_{\mu}^ac^b\right)
+\Omega_{\mu}^a\left(-D^{ab}_{\mu}c^{b}-g\varepsilon^{ab}A_{\mu}^bc\right)
+\Omega_{\mu}\left(-\partial_{\mu}c-g\varepsilon^{ab}A_{\mu}^ac^b\right)\nonumber\\
&&+g\varepsilon^{ab}L^ac^{b}c+\frac{g}{2}\varepsilon^{ab}Lc^{a}c^b
+g\varepsilon ^{ab}F^a\left(\phi\,c^{b}-\phi ^bc\right)-g\varepsilon^{ab}F\phi^ac^b
+\frac{\alpha}{2}\left[b^ab^a-2g\varepsilon^{ab}b^a\bar{c}^bc\right.\nonumber\\
&&\left. +g^2\bar{c}^a\bar{c}^bc^ac^b\right]
+\frac{\beta}{2}\left[g\phi^a\phi^a c^b\bar{c}^b+g\phi^a\phi^bc^a\bar{c}^b
+\phi\phi^a\left(\varepsilon^{ab}b^b-gc\bar{c}^a\right)+g\phi\phi c^a\bar{c}^a\right]\nonumber\\
&&+(\partial_{\mu}\phi^a)(\partial_{\mu}\phi^a)+(\partial_{\mu}\phi)(\partial_{\mu}\phi)
-2g^2\varepsilon^{ab}\left[(\partial_{\mu}\phi)\phi^a A_{\mu}^b
-(\partial_{\mu}\phi^a)\phi A_{\mu}^b+(\partial_{\mu}\phi^a)\phi^b A_{\mu}\right]
\nonumber\\
&&+g^2\left[A_{\mu}^aA_{\mu}^a\left(\phi^b\phi^b+\phi\phi\right)
+A_{\mu}A_{\mu}\phi^a\phi^a -A_{\mu}^aA_{\mu}^b\phi^a\phi^b
-2A_{\mu}^aA_{\mu}\phi^a\phi\right]+\frac{m^2_{\phi}}{2}\left(\phi^a\phi^a+\phi\phi\right)
\nonumber\\
&&\left. \frac{\lambda}{4!}\left[\left(\phi^a\phi^a\right)^2
+2\phi^2\phi^a\phi^a+\phi^4\right] \right\}  \;.
\label{comac}
\end{eqnarray}
Let us  display the quantum numbers of all fields and sources: 
\begin{center}
\begin{tabular}{|l|c|c|c|c|c|}
\hline
\textsc{Fields}$\phantom{\Bigl|}$\! &$A$&$\phi$&$b$&$\bar{c}$&c\\
\hline
\textsc{Dimension}&1&1&2&2&0\\
\textsc{Ghost number}&0&0&0&1&$-1$\\
\textsc{Nature}&B&B&B&F&F\\
\hline
\end{tabular}
\end{center}
\begin{center}
\begin{tabular}{|l|c|c|c|c|c|c|}
\hline
\textsc{Sources}$\phantom{\Bigl|}$\! &$\Omega_{\mu}^a$&$\Omega_{\mu}$&$L^a$&$L$&$F^a$&F\\
\hline
\textsc{Dimension}&3&3&4&4&2&2\\
\textsc{Ghost number}&$-1$&$-1$&$-2$&$-2$&$-1$&$-1$\\
\textsc{Nature}&F&F&B&B&F&F\\
\hline
\end{tabular}
\end{center}
The complete action $\Sigma$ turns out to fulfil a large set of Ward identities, which we enlist below:
\begin{itemize}
\item{The Slavnov-Taylor identity:
\begin{equation}
\mathcal{S}(\Sigma)= 0 \;,  \label{stid}
\end{equation}
with
\begin{equation}
\mathcal{S}(\Sigma)\equiv \int d^{4}x\, \left\{
\frac{\delta\Sigma}{\delta\Omega^{a}_{\mu}}\frac{\delta\Sigma}{\delta A^{a}_{\mu}}
+\frac{\delta\Sigma}{\delta\Omega_{\mu}}\frac{\delta\Sigma}{\delta A_{\mu}}
+\frac{\delta\Sigma}{\delta F^{a}}\frac{\delta\Sigma}{\delta \phi^{a}}
+\frac{\delta\Sigma}{\delta F}\frac{\delta\Sigma}{\delta \phi}
+\frac{\delta\Sigma}{\delta L^{a}}\frac{\delta\Sigma}{\delta c^{a}}
+\frac{\delta\Sigma}{\delta L}\frac{\delta\Sigma}{\delta c}
+b^{a}\frac{\delta\Sigma}{\delta\bar{c}^{a}}   
+b\frac{\delta\Sigma}{\delta\bar{c}}\right\}
\end{equation}
}
Let us also introduce, for further use, the so-called  linearized Slavnov-Taylor operator   $\mathcal{B}_{\Sigma}$, defined as \cite{Piguet:1995er}
\begin{eqnarray}
\mathcal{B}_{\Sigma}&=&\int d^{4}x\,\left\{ 
\frac{\delta\Sigma}{\delta\Omega^{a}_{\mu}}\frac{\delta}{\delta A^{a}_{\mu}}
+\frac{\delta\Sigma}{\delta A^{a}_{\mu}}\frac{\delta}{\delta\Omega^{a}_{\mu}}
+\frac{\delta\Sigma}{\delta\Omega_{\mu}}\frac{\delta}{\delta A_{\mu}}
+\frac{\delta\Sigma}{\delta A_{\mu}}\frac{\delta}{\delta\Omega_{\mu}}
+\frac{\delta\Sigma}{\delta F^{a}}\frac{\delta}{\delta \phi^{a}}
+\frac{\delta\Sigma}{\delta \phi^{a}}\frac{\delta}{\delta F^{a}}
\right. \nonumber\\
&&\left. +\frac{\delta\Sigma}{\delta F}\frac{\delta}{\delta \phi}
+\frac{\delta\Sigma}{\delta \phi}\frac{\delta}{\delta F}
+\frac{\delta\Sigma}{\delta L^{a}}\frac{\delta}{\delta c^{a}}
+\frac{\delta\Sigma}{\delta c^{a}}\frac{\delta}{\delta L^{a}}
+\frac{\delta\Sigma}{\delta L}\frac{\delta}{\delta c}
+\frac{\delta\Sigma}{\delta c}\frac{\delta}{\delta L}
+b^{a}\frac{\delta}{\delta\bar{c}^{a}}
+b\frac{\delta}{\delta\bar{c}}   \right\} 
\end{eqnarray}
The operator $\mathcal{B}_{\Sigma}$ has the important property of being  nilpotent \cite{Piguet:1995er}, {\it i.e.} 
\begin{equation}
\mathcal{B}_{\Sigma} \mathcal{B}_{\Sigma} = 0 \;.
\end{equation}

\item{The diagonal Nakanishi-Lautrup field equation:
\begin{equation}
\frac{\delta\Sigma}{\delta b}=\partial_{\mu}A_{\mu}   \;. \label{db} 
\end{equation}
}
\item{The diagonal anti-ghost equation:
\begin{equation}
\frac{\delta \Sigma}{\delta\bar{c}}+\partial_{\mu}\frac{\delta \Sigma}{\delta\Omega_{\mu}}     = 0    \;. \label{dantigh}
\end{equation}
}
\item{The local diagonal ghost equation \cite{Fazio:2001rm}:
\begin{equation}
\frac{\delta\Sigma}{\delta c}+g\varepsilon^{ab}\bar{c}^a\frac{\delta \Sigma }{\delta b^b}    
=-\partial^2\bar{c}-\partial_{\mu}\Omega_{\mu}
+g\varepsilon^{ab}\left(\Omega_{\mu}^aA_{\mu}^a-L^ac^b+F^a\phi^b\right)   \;. \label{dgh}
\end{equation}
Notice that the right-hand side of eq.\eqref{dgh} is linear in the quantum fields. As such, it is a linear breaking, not affected by the quantum correction \cite{Piguet:1995er}. 
}

\item{The $U(1)$ residual local symmetry:
\begin{equation}
\mathcal{W}^{U(1)}\Sigma=-\partial^2b    \;, \label{u1}
\end{equation}
where
\begin{equation}
\mathcal{W}^{U(1)}\equiv\partial_{\mu}\frac{\delta}{\delta A_{\mu}}+g\varepsilon^{ab} \left\{A^a_{\mu}\frac{\delta}{\delta A^b_{\mu}}+\phi^a\frac{\delta}{\delta\phi^b}+c^a\frac{\delta}{\delta c^b}+\bar{c}^a\frac{\delta}{\delta\bar{c}^b}+b^a\frac{\delta}{\delta b^b}
+\Omega^a_{\mu}\frac{\delta}{\delta\Omega^b_{\mu}}
+F^a\frac{\delta}{\delta F^b}+L^a\frac{\delta}{\delta L^b}
\right\}
\end{equation}
As noticed in \cite{Fazio:2001rm}, the $U(1)$ Ward identity \eqref{u1} can be obtained by anticommuting the diagonal ghost equation, eq.\eqref{dgh}, with the Slavnov-Taylor identity, eq.\eqref{stid}. This identity shows in a very clear way the fact that the diagonal component $A_\mu$ of the gauge field behaves like a $U(1)$ Abelian connection, while all off-diagonal components of the gauge and matter fields play the role of a kind of charged $U(1)$ fields, precisely like in a $QED$-like theory. As already mentioned in the Introduction, this identity expresses  one of the most important characteristic of the maximal Abelian gauge. 
}

\item{The discrete  symmetry
\begin{equation}
\Psi^1\rightarrow\Psi^1\,,\qquad \Psi^2\rightarrow-\Psi^2 \,,\qquad
\Psi^{diag}\rightarrow-\Psi^{diag}  \;, \label{d1} 
\end{equation}
where $\Psi^a$ and $\Psi^{diag}$ stand, respectively,  for all off-diagonal and diagonal fields and sources. As pointed out in \cite{Fazio:2001rm}, this discrete symmetry plays the role of the charge conjugation with respect to the $U(1)$ Cartan subgroup of $SU(2)$. 
}

\item{Finally, looking at the matter sector of the complete action $\Sigma$, we have a second discrete symmetry
\begin{equation}
\phi^a\rightarrow-\phi^a \;, \qquad  \phi \rightarrow-\phi \;, \qquad F^a\rightarrow- F^a \;, \qquad F \rightarrow- F   \;, \label{d2}
\end{equation}
forbidding the appearance of pure matter terms containing odd powers of the scalar fields $(\phi^a, \phi)$. 
}

\end{itemize}

\section{Algebraic characterization of the invariant counterterm and multiplicative renormalizability} 

In order to prove that the complete action $\Sigma$, eq.\eqref{comac}, is multiplicative renormalizable, we follow the algebraic renormalization set up \cite{Piguet:1995er}, and characterise, by means of the Watd identities previously derived,  the most general invariant local counterterm, $\Sigma_{ct}$, which can be freely added to the starting action $\Sigma$. According to the power counting, $\Sigma_{ct}$ is an integrated local polynomial in the fields and external sources of dimension bounded by four and with zero ghost number. Further, we require that the perturbed action, $  (\Sigma +\epsilon \Sigma_{ct})$, satisfies the same Ward identities
and constraints of $\Sigma $ \cite{Piguet:1995er},  to the first order in the perturbation
parameter $\epsilon$, obtaining the following set of constraints: 
\begin{equation} 
\mathcal{B}_{\Sigma}\Sigma_{ct}=0\;, \label{ct1} 
\end{equation}
and 
\begin{eqnarray}
\frac{\delta \Sigma_{ct}}{\delta\bar{c}}+\partial_{\mu}\frac{\delta \Sigma_{ct} }{\delta\Omega_{\mu}}     = 0  \;, \qquad 
\frac{\delta\Sigma_{ct}}{\delta c}+g\varepsilon^{ab}\bar{c}^a\frac{\delta \Sigma_{ct} }{\delta b^b}   =0 \;, \qquad 
\mathcal{W}^{U(1)}\Sigma_{ct}=0\,,\qquad 
\frac{\delta\Sigma_{ct}}{\delta b}=0 \;. \label{ct2}
\end{eqnarray}
The first constraint, eq.\eqref{ct1}, tells us that $\Sigma_{ct}$ belongs to the cohomology of the nilpotent linearized operator $\mathcal{B}_{\Sigma}$ in the space of the integrated local polynomials in the fields and sources bounded by dimension four. From the general results on the BRST cohomolgy of Yang-Mills theories, it follows that  $\Sigma_{ct}$ can be paramterized as follows: 
\begin{equation}
\Sigma_{c.t.}=\Sigma_{0}+   \mathcal{B}_{\Sigma} \Delta^{-1} \;,   \label{pm}
\end{equation}
where $\Sigma_{0}$ stands for the nontrivial part of the cohomolgy of the operator $\mathcal{B}_{\Sigma}$, being given by 
\begin{equation}
\Sigma_{0}=a_0 S_{YM}+\int d^4x\left( a_1\frac{m_{\phi}^2}{2}\phi^A\phi^A
+a_{2} \frac{\lambda}{4!}(\phi^A\phi^A)^2 \right) \;, 
\end{equation}
where $a_0, a_1, a_2$ are free arbitrary coefficients. The second term, $\Delta^{-1}$,  in eq.\eqref{pm}   is  a local integrated polynomial in the fields and sources with dimension four and ghost number $-1$. This term represents the trivial part of the cohomolgy, being parametrized as 
\begin{eqnarray}
\Delta^{-1}&=&\int d^4x \left\{\mathbb{C}^{ab}_4 A^a_{\mu}\Omega^b_{\mu}
+\mathbb{C}_5 A_{\mu}\Omega_{\mu}+\mathbb{C}^{ab}_6 \phi^a F^b
+\mathbb{C}_7 \phi F+\mathbb{C}^{ab}_8 L^ac^b+\mathbb{C}_9 Lc+\mathbb{C}^{ab}_{10} \bar{c}^ab^b 
\right.\nonumber\\
&&  +\mathbb{C}_{11} \bar{c}\,b 
+\mathbb{C}^{ab}_{12}\bar{c}^a\bar{c}^bc+\mathbb{C}^{ab}_{13}\bar{c}^a\bar{c}\,c^b
+\mathbb{C}^{ab}_{14}\phi^a\phi\bar{c}^b
+\mathbb{C}^{ab}_{15}\phi^a\phi^b\bar{c}
+\mathbb{C}^{ab}_{16}A^a_{\mu}A_{\mu}\bar{c}^b+\mathbb{C}^{ab}_{17}A^a_{\mu}A_{\mu}^b\bar{c}
\nonumber\\
&&\left.
+\mathbb{C}^{ab}_{18}m_{\phi}\phi^a\bar{c}^b+\mathbb{C}_{19}m_{\phi}\phi\bar{c}
+\mathbb{C}^{ab}_{21}(\partial_{\mu}A_{\mu}^a)\bar{c}^b
+\mathbb{C}_{21}(\partial_{\mu}A_{\mu})\bar{c}
\right\}   \;, 
\end{eqnarray}
where $\mathbb{C}_{i}, i=4,...,21$ are free parameters. 

After imposition of the conditions \eqref{ct2},  of the discrete symmetries \eqref{d1}, \eqref{d2}, and after a rather lengthy algebraic calculation, we get 
\begin{equation}
\mathbb{C}_5=\mathbb{C}_{9}=\mathbb{C}_{11}=\mathbb{C}_{13}^{ab}=\mathbb{C}_{15}^{ab}=
\mathbb{C}_{17}^{ab}=
\mathbb{C}_{18}^ab=\mathbb{C}_{19}=\mathbb{C}_{21}=0
\end{equation}
and 
\begin{eqnarray}
\mathbb{C}_4^{ab}=\delta^{ab}\mathbb{C}_4\,,\,\,\,
\mathbb{C}_6^{ab}=\delta^{ab}\mathbb{C}_6\,,\,\,\,
\mathbb{C}_7=-\mathbb{C}_6\,,\,\,\,
\mathbb{C}_8^{ab}=\delta^{ab}\mathbb{C}_8\,,\,\,\,
\mathbb{C}_{10}^{ab}=\delta^{ab}\mathbb{C}_{10}\,,\,\,\,\nonumber\\
\mathbb{C}_{12}^{ab}=\varepsilon^{ab}\mathbb{C}_{12}=-\varepsilon^{ab}\mathbb{C}_{10}\,,\,\,\,
\mathbb{C}_{14}^{ab}=\varepsilon^{ab}\mathbb{C}_{14}\,,\,\,\,
\mathbb{C}_{16}^{ab}=\varepsilon^{ab}\mathbb{C}_{16}\,,\,\,\,
\mathbb{C}_{20}^{ab}=\delta^{ab}\mathbb{C}_{20}=-\delta^{ab}\mathbb{C}_{16}\,,\,\,\,
\end{eqnarray}
Therefore, for the final expression of the  most general  counterterm $\Sigma_{ct}$, we obtain 
\begin{eqnarray}
\Sigma_{ct} &=&\int d^4x \left( \frac{a_0}{4}\left(F_{\mu\nu}^aF_{\mu\nu}^a+F_{\mu\nu}F_{\mu\nu}\right)
+a_1\frac{m_{\phi}^2}{2}\phi^A\phi^A+a_{2}\frac{\lambda}{4!}(\phi^A\phi^A)^2  \right) \nonumber \\
& +& \mathcal{B}_{\Sigma} \int d^4x \left[
\mathbb{C}_4 A^a_{\mu}\Omega^a_{\mu}+\mathbb{C}_6\left(\phi^a F^a-\phi F\right)
+\mathbb{C}_8 L^ac^a
+\mathbb{C}_{10}\alpha \left(\bar{c}^ab^a-\varepsilon^{ab}\bar{c}^a\bar{c}^b c \right) \right]  \nonumber \\
&  + & \mathcal{B}_{\Sigma} \int d^4x \left[  \mathbb{C}_{14}\beta\varepsilon^{ab} \phi^a \phi\bar{c}^b
+\mathbb{C}_{16}\bar{c}^aD_{\mu}^{ab}A^b_{\mu} \right] \nonumber\\
&=&\int d^4x\left( \frac{a_0}{4}\left(F_{\mu\nu}^aF_{\mu\nu}^a+F_{\mu\nu}F_{\mu\nu}\right)
+a_1\frac{m_{\phi}^2}{2}\phi^A\phi^A+a_{2}\frac{\lambda}{4!}(\phi^A\phi^A)^2+
\mathbb{C}_{4}\left[ \frac{\delta S_{YM}}{\delta A^a_{\mu}}A^a_{\mu}+b^aD_{\mu}^{ab}A_{\mu}^b
\right.\right.\nonumber\\
&& +g\varepsilon^{ab}\left(\bar{c}^acD_{\mu}^{bc}A_{\mu}^c
-\Omega_{\mu}A_{\mu}c^b+\bar{c}\partial_{\mu}(A_{\mu}^ac^b)\right)
+2g^2\left(\bar{c}^ac^a+\phi^a\phi^a+\phi\phi\right)A_{\mu}^bA_{\mu}^b
\nonumber\\
&&\left.-2g^2\left(\bar{c}^ac^b+\phi^a\phi^b\right)A_{\mu}^aA_{\mu}^b
+2g\varepsilon^{ab}A_{\mu}^a\left((\partial_{\mu}\phi)\phi^b-(\partial_{\mu}\phi^b)\phi\right)
\right]+\mathbb{C}_6\left[2(\partial_{\mu}\phi^a)(\partial_{\mu}\phi^a) \right.
\nonumber\\
&& -2(\partial_{\mu}\phi)(\partial_{\mu}\phi)
-4g^2\varepsilon^{ab}(\partial_{\mu}\phi^a)\phi A_{\mu}^b
+2g^2A_{\mu}^aA_{\mu}^a\left(\phi^a\phi^a-\phi\phi\right)
+2g^2\left(A_{\mu}A_{\mu}\phi^a\phi^a-A_{\mu}^aA_{\mu}^b\phi^a\phi^b\right)\nonumber\\
&& \left. +m_{\phi}^2\left(\phi^a\phi^a-\phi\phi\right)
+\frac{\lambda}{3!}\left((\phi^a\phi^a)^2-\phi^4\right)
+\beta g\left(\phi^a\phi^ac^b\bar{c}^b+\phi^a\phi^bc^a\bar{c}^b-\phi\phi c^a\bar{c}^a\right)
\right]\nonumber\\
&& + \mathbb{C}_8\left[-\bar{c}^a\partial^2c^a
+2g\varepsilon^{ab}\bar{c}^aA_{\mu}\partial_{\mu}c^b
+g^2\bar{c}^ac^a\left(A_{\mu}A_{\mu}-A_{\mu}^bA_{\mu}^b\right)
+g^2\bar{c}^ac^bA_{\mu}^aA_{\mu}^b-g\varepsilon^{ab}Lc^ac^b \right.\nonumber\\
&& +\Omega^a_{\mu}D_{\mu}^{ab}c^b+g\varepsilon^{ab}\Omega_{\mu}c^aA_{\mu}^b
-g\varepsilon^{ab}F^a\phi\,c^b+g\varepsilon^{ab}F\phi^ac^b
-\alpha g^2\bar{c}^ac^a\bar{c}^bc^b+\frac{\beta}{2} g\bar{c}^ac^a\left(\phi^b\phi^b+\phi^2\right)
\nonumber\\
&&\left. \frac{\beta}{2}\phi^a\phi^b\bar{c}^bc^a \right]
+\mathbb{C}_{10}\alpha\left[b^ab^a-2g\varepsilon^{ab}b^a\bar{c}^bc+g^2\bar{c}^ac^a\bar{c}^bc^b
\right]+\mathbb{C}_{14}\beta\left[\phi\phi^a\left(\varepsilon^{ab}b^b+gc\bar{c}^a\right)
\right.\nonumber\\
&&\left. gc^a\bar{c}^a\left(\phi^a\phi^a+\phi^2\right)+g\phi^a\phi^ac^b\bar{c}^a\right]
+\mathbb{C}_{16}\left[\bar{c}^a\partial^2c^a
-2g\varepsilon^{ab}\bar{c}^aA_{\mu}\partial_{\mu}c^b-g^2\bar{c}^ac^bA_{\mu}^aA_{\mu}^b
\right.\nonumber\\
&&\left. +g^2\bar{c}^ac^a\left(A_{\mu}^bA_{\mu}^b+A_{\mu}A_{\mu}\right)
+2\varepsilon^{ab}\bar{c}^ac\,D_{\mu}^{bc}A_{\mu}^c+b^aD_{\mu}^{ab}A_{\mu}^b
\right]
\left. \right)      \;. \label{mgct}
\end{eqnarray}

\subsection{Renormalization factors}

After having identified the most general counterterm, expression \eqref{mgct}, it remains to check if it can be reabsorbed through a multiplicative redefinition of the fields, sources, coupling constant and parameters of the starting action, according to 
\begin{equation}
\Sigma(\Psi_0,\Gamma_0,\xi_0)=\Sigma(\Psi,\Gamma,\xi)+\epsilon\Sigma_{c.t.}(\Psi,\Gamma,\xi)  + O(\epsilon^2) \;, 
\label{sct}
\end{equation} 
where 
\begin{eqnarray}
\Psi &=&\{A^a_{\mu},A_{\mu},\phi^a,\phi,b^a,b,\bar{c}^a,c^a\}   \;, \nonumber\\
\Gamma &=&\{\Omega_{\mu}^a,\Omega_{\mu},F^a,F,L^a,L,\}  \;, \nonumber\\
\xi &=&\{g,m_{\phi},\lambda,\alpha,\beta\} \;, 
\end{eqnarray}
and the so-called bare quantities $(\Psi_0,\Gamma_0,\xi_0)$ are defined as

\begin{equation}
\Psi_0=Z_{\Psi}^{1/2}\Psi\,,\qquad\Gamma_0=Z_{\Gamma}\Gamma\,,\qquad\xi_0=Z_{\xi}\xi
\end{equation}
By direct inspection  of equation  \eqref{sct}, for the renormalization factors we obtain 
\begin{eqnarray}
Z_A^{1/2}&=&1+\epsilon(2a_0+\mathbb{C}_4)\\
(Z_A^{diag})^{1/2}&=&1+2\epsilon a_0\\
Z_b^{1/2}&=&1+\epsilon(-2a_0+\mathbb{C}_{16})\\
(Z_b^{diag})^{1/2}&=&1-2\epsilon a_0\\
Z_c^{1/2}&=&1-\epsilon\mathbb{C}_8\\
Z_{\bar{c}}^{1/2}&=&1+\epsilon\mathbb{C}_{16}\\
Z_{\phi}^{1/2}&=&1+\epsilon\mathbb{C}_6\\
(Z_{\phi}^{diag})^{1/2}&=&1-\epsilon\mathbb{C}_6\\
Z_g&=&1-2\epsilon a_0\\
Z_{m_{\phi}}&=&1+\frac{\epsilon}{2}a_1\\
Z_{\alpha}&=&1+2\epsilon\left(2a_0+\mathbb{C}_{10}-\mathbb{C}_{16}\right)\\
Z_{\beta}&=&1+\epsilon\left(-2a_0+2\mathbb{C}_{14}+\mathbb{C}_{16}\right)\\
Z_{\lambda}&=&1+\epsilon a_2 \;.  \label{ren} 
\end{eqnarray}
It is worth noticing that the diagonal ghosts do not need to be renormalized, a property which follows directly from the diagonal ghost equation  \eqref{dgh}. This concludes the algebraic proof of the all orders multiplicative renormalizability of the action $\Sigma$, eq.\eqref{comac}. Finally, we note that the non-renormalization theorem of the maximal Abelian gauge \cite{Fazio:2001rm}
\begin{equation}
Z_g(Z_A^{diag})^{1/2}=1  \;, \label{nren}
\end{equation}
remains true in the presence of matter fields.

\section{Conclusion}

In this work we have addressed the issue of the renormalization of Yang-Mills theories in the maximal Abelian gauge in the presence of scalar matter fields. Our main observation is that, due to the non-linearity of the gauge fixing condition, eq.\eqref{offgauge}, a new quartic interaction term between scalar matter fields and off-diagonal Faddeev-Popov ghosts is required for renormalizabilty.  Moreover, this new quartic interaction turns out to be described by an exact BRST invariant term, as expressed by eq.\eqref{sbeta}, a feature which ensures that the final gauge fixed action, eq.\eqref{comac}, is BRST invariant and multiplicative renormalizable to all orders, as proven in Sect.5. \\\\Although the proof of the renormalizability given here refers to the gauge group $SU(2)$, it can be easily generalised to other gauge groups as well as to other representations of the scalar fields. The inclusion of the usual Dirac action for spinors does not pose any additional problem. Also, unlike the case of scalar matter fields, BRST invariance and power counting do not allow for additional interaction terms between spinors and Faddeev-Popov ghosts.\\\\The analysis of the all orders perturbative renormalizability of the maximal Abelian gauge in presence of matter fields is the first necessary step towards the investigation of the non-perturbative effects of the Gribov copies, which deeply affect the maximal Abelian gauge \cite{Capri:2006vv,Capri:2006cz,Capri:2008ak,Capri:2008vk,Capri:2010an}. The study of this issue in presence of matter fields  is currently under investigation \cite{prep}.

\section*{Acknowledgments}
The Conselho Nacional de Desenvolvimento Cient\'{\i}fico e
Tecnol\'{o}gico (CNPq-Brazil), the Faperj, Funda{\c{c}}{\~{a}}o de
Amparo {\`{a}} Pesquisa do Estado do Rio de Janeiro,  the
Coordena{\c{c}}{\~{a}}o de Aperfei{\c{c}}oamento de Pessoal de
N{\'{\i}}vel Superior (CAPES),  are gratefully acknowledged.

\end{document}